# Atomic-scale confinement of optical fields

Johannes Kern[1], Swen Großmann[1], Nadezda V. Tarakina[2,3], Tim Häckel[1], Monika Emmerling[2], Martin Kamp[2], Jer-Shing Huang[4], Paolo Biagioni[5], Jord C. Prangsma[1] & Bert Hecht[1,†]

[1]Nano-Optics & Biophotonics Group, Experimentelle Physik 5, Physikalisches Institut, Wilhelm-Conrad-Röntgen-Center for Complex Material Systems, Universität Würzburg, Am Hubland, D-97074 Würzburg, Germany

[2]Technische Physik, Physikalisches Institut, Wilhelm-Conrad-Röntgen-Center for Complex Material Systems, Universität Würzburg, Am Hubland, D-97074 Würzburg, Germany

[3]Experimentelle Physik 3, Physikalisches Institut, Wilhelm-Conrad-Röntgen-Center for Complex Material Systems, Universität Würzburg, Am Hubland, D-97074 Würzburg, Germany

[4]Department of Chemistry, National Tsing Hua University, Hsinchu 30013, Taiwan

[5]CNISM - Dipartimento di Fisica, Politecnico di Milano, Piazza Leonardo da Vinci 32, 20133 Milano, Italy

[†]to whom correspondence should be addressed (hecht@physik.uni-wuerzburg.de)

**In the presence of matter there is no fundamental limit preventing confinement of visible light even down to atomic scales. Achieving such confinement and the corresponding intensity enhancement inevitably requires simultaneous control over atomic-scale details of material structures and over the optical modes that such structures support. By means of self-assembly we have obtained side-by-side aligned gold nanorod dimers with robust atomically-defined gaps reaching below 0.5 nm. The existence of atomically-confined light fields in these gaps is demonstrated by observing extreme Coulomb splitting of corresponding symmetric and anti-symmetric dimer eigenmodes of more than 800 meV in white-light scattering experiments. Our results open new perspectives for atomically-resolved spectroscopic imaging, deeply nonlinear optics, ultra-sensing, cavity optomechanics as well as for the realization of novel quantum-optical devices.**

The interaction of light and matter, i.e. absorption and emission of photons, can be considerably enhanced in the presence of strongly localized and therefore highly intense optical near fields[1]. Plasmonic antennas consisting of pairs of closely spaced metal nano particles have gained much attention in this context since they provide the possibility to strongly concentrate optical fields into the gap between the two metal particles[2-4]. Pairs of closely spaced metal nanoparticles supporting plasmonic gap resonances consequently find broad applications, e.g. in single-emitter surface-enhanced spectroscopy[5, 6], quantum optics[7], extreme nonlinear optics[8-10], optical trapping[11], metamaterials[12, 13] and molecular opto-electronics[14].

The success of metal-insulator-metal structures is based on two fundamental properties of their anti-symmetric electromagnetic gap modes. (i) As a direct consequence of the boundary conditions, the dominating field components normal to the metal-dielectric interfaces are sizable only inside the dielectric gap. This means that for anti-symmetric gap modes the achievable field confinement is not limited by the skin depth of the metal, but is solely determined by the actual size of the gap. (ii) Since the free electrons of the metal respond resonantly to an external optical frequency field, enormous surface charge accumulations, accompanied by ultra-intense optical near fields, will occur. In addition, with decreasing gap width, stronger attractive coulomb forces[4] across the gap lead to further surface-charge accumulation and a concomitantly increased near-field intensity enhancement. We therefore conclude that an experimental realization of atomic-scale concentration of electromagnetic fields at visible frequencies is possible but it requires atomic-scale shape control of the field-confining structure, i.e. the gap, as well as a careful assignment and selection of suitable optical modes.

Here we achieve atomic-scale confinement of electromagnetic fields at visible frequencies by combining for the first time both atomic-scale shape control of the field confining structure[15] as well as a careful selection and assignment of suitable optical modes[16-18]. We study single-crystalline nanorods which self-assemble into side-by-side aligned dimers with gap widths below 0.5 nm. Side-by-side aligned nanorod dimers possess various distinguishable symmetric and anti-symmetric modes in the visible range[19]. In contrast to previous work[20-22] we demonstrate full control over symmetric and anti-symmetric optical modes by means of white-light scattering experiments. We experimentally demonstrate the presence of atomic-scale light confinement in these structures by observing an extreme > 800 meV hybridization splitting of corresponding symmetric and anti-symmetric dimer modes. Our results open new perspectives for atomically-resolved spectroscopic imaging, deeply nonlinear optics and attosecond physics, cavity optomechanics and ultra-sensing as well as quantum optics.

To obtain nanostructures with very small gaps, we dropcast chemically grown nanorods onto an ITO-coated glass cover slip. The nanorods exhibit an intrinsic size distribution with lengths and diameters ranging from 62 to 78 nm and 25 to 32 nm, respectively (Supplementary Fig. S1). Gold nanorods tend to align side-by-side due to the joint action of capillary forces[23] and the interdigitation of the nanorod's surfactant layers in order to minimize the hydrophilic-hydrophobic interaction in presence of water[24, 25]. To characterize the typical gap width of self-assembled gold nanorod dimers, high-angle annular dark field scanning transmission electron microscopy (HAADF-STEM) was performed (Methods). Careful analysis yields gaps ranging from 0.5 to ~2 nm with an average gap width of 1.3 nm (Supplementary Fig. S2). Figure 1 shows a HAADF-STEM image of a typical nanorod dimer and a zoom to the gap region, revealing the atomic structure of the left rod.

In order to facilitate the assignment of nanorod dimer plasmon resonances we have performed numerical simulations for a range of gap widths using the Finite-Difference-Time-Domain (FDTD) method (Methods). Variation of the gap width strongly influences the electromagnetic coupling strength and therefore causes characteristic shifts of the modes[4, 17, 18]. For all investigated gap widths (0.3 to 3.6 nm) we observe four characteristic resonances in the visible spectral range. The coupled transverse mode (Fig. 2a) and the longitudinal mode (Fig. 2b) both are dipole modes of the coupled system. The remaining two resonances (Fig. 2c,d) are assigned to the anti-symmetric hybridized counter part of the longitudinal mode as well as the 2$^{nd}$-order anti-symmetric mode. These latter modes can also be interpreted as the 1$^{st}$- and 2$^{nd}$-order mode of a 'cavity' that consists of two rods supporting a highly-localized metal-insulator-metal (MIM) mode[26, 27] propagating along the gap and being reflected at the rod ends[28]. Cavity resonances occur for extended dimer lengths of $L=m\cdot\lambda/(2\cdot n_{eff})$, where m is the order of the resonance[29]. Note that in our structures, due to the narrow gap, the effective index $n_{eff}$ can be as high as 14. The resulting strong localization of the optical near-field in all three dimensions leads to unprecedentedly small modal volumes[30] of 463 nm$^3$ for the 2$^{nd}$-order cavity mode.

In view of experimental studies, the polarization-dependent radiation patterns of the dimer eigenmodes need to be determined to enable the assignment of observed peaks as well as to ensure optimal excitation[31]. The far-field polarization of each mode is generally oriented parallel to the symmetric plane of a mode. For the coupled longitudinal/transverse mode such a plane obviously exists in the longitudinal/transverse direction, respectively. Accordingly, the 2$^{nd}$-order cavity mode emission exhibits transverse polarization. The 1$^{st}$-order cavity mode has a quadrupolar character with two anti-symmetric planes and emission to the far-field is therefore expected to be weak and is found to be polarized along the longitudinal direction. The corresponding simulated emission patterns are plotted in Fig. S3 (Supplementary).

Optical characterization of nanorod dimers was performed using an asymmetric dark-field white-light scattering scheme employing an off-axis needle-like beam to enable efficient excitation of dark and bright modes of all symmetries[32]. Even the robust side-by-side self-assembly leads to dimers with slightly variable geometries as well as slight variations in gap width. For our data analysis we therefore only selected the most symmetric structures by means of scanning electron microscopy (Supplementary). Typical scattering spectra of such dimers (Fig. 2e) reveal four resonances within the observation window (500 to 1000 nm wavelength). Two resonances exhibit an emission polarization along the longitudinal axis (blue disk, orange star) of the nanorod dimer, while the other two resonances show transverse polarization (green triangle, red square). The resonance wavelength and the quality factor of each resonance are obtained by fitting a Lorentzian lineshape to each peak. Taking resonance position and far-field polarization into account the observed resonances can unambiguously be identified as the coupled transverse (green triangle), coupled longitudinal (blue disk), 1$^{st}$-order (orange star) and 2$^{nd}$-order (red square) cavity mode by comparing to FDTD simulations.

The excitation efficiency of each resonance depends on the orientation of the structure with respect to the excitation plane and polarization[31]. Indeed the scattering intensity of each resonance can be tuned by varying the orientation of the nanorod dimer with respect to the excitation plane (data not shown). For all investigated structures, both cavity resonances and the longitudinal resonance could be easily identified according to their emission polarization and their excitation pattern. The scattering intensity of the coupled transverse mode is typically very weak, since it falls into a region, where gold exhibits strong interband damping. Therefore, the weak transverse resonance could not always be identified with sufficient fidelity.

For further data analysis we exploit the fact, that the energy splitting (ΔE) of the hybridized pair, longitudinal and 1$^{st}$-order cavity resonance, is a measure of the coupling strength which, due to the strong confinement, depends on the gap width and the refractive index of material inside the gap. In Fig. 2f we plot energy splitting versus resonance position for simulated dimers with different gap width and refractive index in the gap (solid lines) along with the experimentally obtained values of all investigated dimers (symbols). We find good agreement between experiment and simulation, e.g. the characteristic crossing point between longitudinal and 2$^{nd}$-order cavity mode, indicating that variations in the gap are indeed the dominant source of spectral shifts in our experiments.

As a next step we study the simulated dependence of energy splitting and gap width for the extreme cases of no material (n=1) and dense organic material (n=1.3) in the gap. An approximately linear relationship between inverse energy splitting and gap width is found in the considered gap width range of 0.3 to 3.6 nm (inset in Fig. 2f). Each nanorod dimer therefore possesses an internal ruler that sensitively reports the gap width in a range of few up to tens of ångströms once calibrated by FDTD simulations.

Assuming a refractive index of n=1.15 in the gap, we are now able to assign a gap width to each nanorod dimer that was investigated (Fig. 2f). The resulting gap widths range from ~2 to <0.3 nm, in good agreement with our HAADF-STEM studies. Two dimers show extreme mode splittings of up to 800 meV, falling outside of our calibration regime. The gap width for these two dimers is therefore smaller than 0.3 nm, about the size of a gold atom. It is interesting to note that systematic deviations between simulations and experimental data at small gap widths seem to be most pronounced for anti-symmetric modes. Some amount of scatter in the experimental data can be attributed to small remaining variations in geometrical parameters, such as rod length, shape of the end caps and slight asymmetries of the dimers (Supplementary Fig. S4), as well as in the dielectric function of the surrounding and the gold itself. These uncertainties make it difficult to judge if our experiments can still be described by local Maxwell theory or if quantum effects like electron tunneling have a significant influence for gaps <0.5 nm as suggested by recent theoretical work[33].

Analysis of the quality-factor of each resonance was also performed (Supplementary Fig. S4) and yielded good agreement between simulation and experiments for the coupled longitudinal resonance. However, for both observed cavity resonances the experimentally measured quality factor is significantly lower than in simulations, even when including the effect of structural asymmetries that may lead to additional radiative losses of dark modes. As a matter of fact, unexpectedly low quality factors of experimentally observed dark modes are a common feature in the literature[32, 34]. Since single-crystalline structures are used in the present study we exclude effects of domain boundaries as a major source of the observed additional broadening, leaving non-local effects, electron tunneling and surface-related effects[35, 36], all not considered in simulations, as probable causes of this behavior.

The atomic-scale confinement and enhancement of visible light for the cavity modes (Fig. 2c and d) as well as their spectral tunability and large sensitivity to the gap width will have applications in various fields of research. In order to realize optical spectroscopy with near-atomic resolution, e.g. exploiting Raman scattering, dimers with flat end facets could be embedded in suitable tip structures for scanning probe microscopy. The extremely strong dependence of the resonance wavelength on the gap will open new possibilities in the field of cavity optomechanics[37]. It is also conceivable that quantum-optical effects, like the so-called photon blockade[38, 39], may become observable in gold nanorod dimer resonators due to their extremely small modal volumes. Furthermore, generation of high harmonics has been predicted to occur in metal particle dimers featuring ultrasmall gaps[33] which could give rise to novel solid-state sources of attosecond laser pulses[40].

**Methods**

**Sample preparation**

Gold nanorods (NanorodzTM 30-25-650) were purchased from Nanopartz (Loveland, USA). The nanorods were dropcasted on an ITO-coated coverglass (Menzel Gläser, Braunschweig, Germany), on which a gold marker structure was prefabricated by electron beam lithography. The marker structures are designed to be easily observable in both dark-field optical microscopy and SEM (Helios Nanolab, FEI Company), allowing for a precise and fast identification of individual nanostructures as well as recording of scattering spectra. Due to the coffee stain effect the nanorods gather at the rim of the droplet and isolated dimers can be quickly identified in a Scanning Electron Microscope (SEM). To remove excess Cetyl-Trimethyl-Ammoniumbromid (CTAB), which is the surfactant stabilizing the nanorods, the sample was rinsed in 70° warm ultra-pure water (Milli-Q) and cleaned with an oxygen plasma. Further it has been observed that high voltages (25 kV), rather low magnifaction (25k-40k) and short dwell times (3 µs) lead to a minimal carbon deposition during SEM imaging.

**STEM**

High-angle annular dark field scanning transmission electron microscopy (HAADF-STEM) experiments were performed using a FEI Titan 80-300 electron microscope operating at 300 kV and having a STEM resolution of 0.13 nm. Gold nanorods have been deposited on a holey carbon film. For each pair of rods a tilt series in the direction perpendicular to their long facets was performed, the images displaying the widest gaps were taken for further analysis. The gap width was determined for twelve pairs of parallelly aligned rods both by visual inspection and by a least square fitting of the transmission profile of the gap assuming a circular cross section of the nanorods. Both methods yielded an average gap width of 1.3 nm. Ten out of twelve structures had gaps between 0.5 and ~2 nm, one pair had a gap width too small to be determined and one outlier had a gap width of 3.5 nm. Further details of gap determination can be found in the supplementary information. For HAADF-STEM images with atomic columns resolved, pairs of rods were oriented along the [110] zone-axis.

**Numerical Simulations**

Numerical simulations were performed using commercial FDTD software (FDTD Solutions v6.5.11, Lumerical Solutions). The rods were modeled as cylindrical particles with spherical endcaps. The diameter of the rods was set to 30 nm and the length was set to 70 nm, in accordance with results from electron microscopy. The gap width between two side-by-side aligned rods was varied from 0.3nm to 3.6 nm. The mesh was chosen to include 10 mesh cells within the gap and the minimal mesh size is set to 0.1 nm³. It is worth noting that the FDTD algorithm has proven to give robust results for resonance position and quality factor even for a comparatively small number of mesh cells in the gap (Supplementary). To study the effect of refractive index, a dielectric material was added to the gap region with refractive index of n=1 or n=1.3. The substrate consists of a 200 nm thick ITO layer (dielectric material with n=1.7) on top of a thick SiO2 layer (dielectric material with n=1.455). The dielectric constant of gold is modeled by an analytical fit to experimental data[41]. The eigenmodes of the structures are determined by choosing an efficient excitation geometry and applying apodization in the time domain to remove influences of the source.

**Dark-Field Microscopy**

To obtain white-light scattering spectra, an off-axis needle-like beam (1–2 mm diameter collimated beam) originating from a halogen lamp (Axiovert, Zeiss) is focused by a microscope objective (Plan-Apochromat, 63, N.A. = 1.4, Zeiss) to the sample plane. The off-axis beam is parallel to, but displaced from the objective's optical axis, such that it hits the sample surface at an angle larger than the critical angle of total internal reflection. It therefore undergoes total internal reflection at the sample plane. Illuminated nanorods scatter light into a broad angular range collected by the same objective while the reflected excitation beam is blocked by a small beam stop. An analyser (LPVIS, Thorlabs) is used to select the polarization of the scattered light before the entrance slit (200 μm) of the spectrometer (ACTON SpectraPro 2300i, 150 grooves per mm grating blazing at 500 nm). The typical acquisition time of the charge-coupled device (Andor DV434-BV) is 20s.

**References**

1. Novotny, L., Hecht, B. *Principles of Nano-Optics.* (Cambridge University Press, Cambridge, 2006).

2. Mühlschlegel P., Eisler H.-J., Martin O. J. F., Hecht B., Pohl D. W. Resonant Optical Antennas. *Science*, **308**, 1607-1609 (2005)

3. Schuck P. J., Fromm D. P., Sundaramurthy A., Kino G. S., Moerner W. E. Improving the Mismatch between Light and Nanoscale Objects with Gold Bowtie Nanoantennas. *Phys. Rev. Lett.* **94,** 017402 (2005)

4. Biagioni, P., Huang, J.-S., Hecht, B. Nanoantennas for visible and infrared radiation. *Rep. Prog. Phys.* **75**, 024402 (2012).

5. Xu, H., Bjerneld E. J., Käll M., Börjesson L. Spectroscopy of Single Hemoglobin Molecules by Surface Enhanced Raman Scattering. *Phys. Rev. Lett.* **83**, 4357-4360 (1999).

6. Kinkhabwala, A. *et al.* Large single-molecule fluorescence enhancements produced by a bowtie nanoantenna. *Nature Photon.* **3**, 654-657 (2009).

7. Schietinger, S., Barth, M., Aichele, T., Benson, O. Plasmon-Enhanced Single Photon Emission from a Nanoassembled Metal-Diamond Hybrid Structure at Room Temperature. *Nano Lett.* **9**, 1694-1698 (2009).

8. Palomba, S., Novotny, L. Near-Field Imaging with a Localized Nonlinear Light Source. *Nano Lett.* **9**, 3801-3804 (2009).

9. Kim, S. *et al.* High-harmonic generation by resonant plasmon field enhancement. *Nature* **453**, 757-760 (2008).

10. Hanke, T. *et al.* Efficient Nonlinear Light Emission of Single Gold Optical Antennas Driven by Few-Cycle Near-Infrared Pulses. *Phys. Rev. Lett.* **103**, 257404 (2009).

11. Juan, M. L., Righini, M., Quidant, R. Plasmon nano-optical tweezers. *Nature Photon*. **5**, 349-356 (2011).

12. Fan, J. A. *et al.* Self-assembled Plasmonic Nanoparticle Clusters. *Science* **328**, 1135-1138 (2010).

13  Liu, N. *et al.* Three-dimensional photonic metamaterials at optical frequencies. *Nature Materials* **7**, 31-37 (2008).

14  González, M. T. *et al.* Electrical Conductance of Molecular Junctions by a Robust Statistical Analysis. *Nano Lett.* **6**, 2238-2242 (2006).

15  Katz-Boon, H. *et al.* Three-Dimensional Morphology and Crystallography of Gold Nanorods. *Nano Lett.* **11**, 273-278 (2011).

16  Huang, J.-S. *et al.* Mode Imaging and Selection in Strongly Coupled Nanoantennas. *Nano Lett*. **10**, 2105-2110 (2010).

17  Atay, T., Song, J.-H., Nurmikko, A. V. Strongly Interacting Plasmon Nanoparticle Pairs: From Dipole-Dipole Interaction to Conductively Coupled Regime. *Nano Lett*. **4**, 1627-1631 (2004).

18  Lassiter, J. B. *et al.* Close Encounters between Two Nanoshells. *Nano Lett*. **8**, 1212-1218 (2008).

19  Lyvers, D. P., Moon, J.-M., Kildishev, A. V., Shalaev, V. M., Wei, A. Gold Nanorod Arrays as Plasmonic Cavity Resonators. *ACS Nano* **2**, 2569-2576 (2008).

20  Funston, A. M., Novo, C., Davis, T. J., Mulvaney, P. Plasmon Coupling of Gold Nanorods at Short Distances and in Different Geometries. *Nano Lett*. **9**, 1651-1658 (2009).

21  Slaughter, L. S., Wu, Y., Willingham, B. A., Nordlander, P., Link, S. Effects of Symmetry Breaking and Conductive Contact on the Plasmon Coupling in Gold Nanorod Dimers. *ACS Nano* **4**, 4657-4666 (2010).

22  Marhaba, S. *et al.* Surface Plasmon Resonance of Single Gold Nanodimers near the Conductive Contact limit. *J. Phys. Chem. C.* **113**, 4349-4356 (2009).

23  Nikoobakht, B., Wang, Z. L., El-Sayed, M. A. Self-Assembly of Gold Nanorods. *J. Phys. Chem. B.* **104**, 8635-8640 (2000).

24  Jana, N. R. *et al.* Liquid crystalline assemblies of ordered gold nanorods. *J. Mater. Chem.* **12**, 2909-2912 (2002).

25  Kawamura, G., Yang, Y., Nogami, M. Facile assembling of gold nanorods with large aspect ratio and their surface-enhanced Raman scattering properties. *Appl. Phys. Lett*. **90**, 261908 (2007).

26  Miyazaki, H. T., Kurokawa, Y. Squeezing Light Waves into a 3-nm-Thick and 55-nm-Long Plasmon Cavity. *Phys. Rev. Lett.*  **96**, 097401 (2006).

27  Dionne, J. A., Lezec, H. J., Atwater, H. A. Highly Confined Photon Transport in Subwavelength Metallic Slot Waveguides. *Nano Lett*. **6**, 1928-1932 (2006).

28  Huang, J.-S., Feichtner, T., Biagioni, P., Hecht, B. Impedance Matching and Emission Properties of Nanoantennas in an Optical Nanocircuit. *Nano Lett*. **9**, 1897-1902 (2009).

29  Taminiau, T. H. , Stefani, F. D., van Hulst, N. F. Optical Nanorod Antennas Modeled as Cavities for Dipolar Emitters: Evolution of Sub- and Super-Radiant Modes. *Nano Lett.* **11**, 1020-1024 (2011).

30  Koenderink, A. F. On the use of Purcell factors for plasmon antennas. *Opt. Lett.* **35**, 4208-4210 (2010).

31  Dorfmüller, J. *et al.* Fabry-Pérot Resonances in One-Dimensional Plasmonic Nanostructures. *Nano Lett*. **9**, 2372-2377 (2009).

32  Yang, S. C. *et al.* Plasmon Hybridization in Individual Gold Nanocrystal Dimers: Direct Observation of Bright and Dark Modes. *Nano Lett*. **10**, 632-637 (2010).

33  Marinica, D. C. , Karzansky, A. K., Nordlander, P., Aizpurua, J., Borisov, A. G. Quantum Plasmonics: Nonlinear Effects in the Field Enhancement of a Plasmonic Nanoparticle Dimer. *Nano Lett.* **12**, 1333-1339 (2012).

34  Vesseur, E. J. R., Garcia de Abajo, F. J., Polman, A. Modal Decomposition of Surface-Plasmon Whispering Gallery Resonators. *Nano Lett.* **9**, 3147-3150 (2009).

35  Hövel, H., Fritz, S., Hilger, A., Kreibig, U., Vollmer, M. Width of cluster plasmon resonances: Bulk dielectric functions and chemical interface damping. *Phys. Rev.  B* **48**, 18178-18188 (1993).

36  Lermé, J. *et al.* Size Dependence of the Surface Plasmon Resonance Damping in Metal Nanospheres. *J.  Phys. Chem. Lett* **1**, 2922-2928 (2010).

37  Kippenberg, T. J., Vahala, K. J. Cavity Optomechanics: Back-Action at the Mesoscale. *Science* **321**, 1172-1176 (2008).

38  Imamoglu, A., Schmidt, H., Woods, G., Deutsch, M. Strongly interacting photons in a nonlinear cavity. *Phys. Rev. Lett.* **79**, 1467-1470 (1997).

39  Smolyaninov, I. I., Zayats, A. V., Gungor, A., Davis, C. C. Single-photon tunneling via localized surface plasmons. *Phys. Rev. Lett.* **88**, 187402 (2002).

40  Brabec, T., Krausz, F. Intense few-cycle laser fields: Frontiers of nonlinear optics. *Rev. Mod. Phys.* **72**, 545-591 (2000).

41  Etchegoin, P. G., Le Ru, E. C., Meyer, M. An analytic model for the optical properties of gold. *J. Chem. Phys.* **125**, 164705 (2006).

**Acknowledgement**

The authors thank James Pond (Lumerical Inc.) and A. Guerrero-Martinez for valuable discussions. Financial support of the VW-foundation (grant I/84036) and the DFG (HE 5618/1-1) is gratefully acknowledged.

**Author contributions**

J.K., B.H., P.B., and J.-S. H. conceived the idea. J.K., S.G., T.H. and M.E. prepared the samples. N.V.T, M.K. and J.C.P. performed STEM experiments and analysis. J.K., S.G. and T.H. carried out the optical experiments. J.K., J.C.P and J.-S.H. performed SEM. J.K. designed and implemented numerical simulations. J.K., S.G. and J.C.P analysed the data. J.K., S.G., J.P. and B.H. wrote the manuscript. All authors contributed to scientific discussion and critical revision of the article. P.B., J.C.P. and B.H. supervised the study.

**Competing financial interests**

The authors declare no competing financial interests.

**Additional information**

Supplementary information accompanies this paper at www.nature.com. Reprints and permission information is available online at http://npg.nature.com/reprintsandpermissions/. Correspondence and requests for materials should be addressed to B.H. (hecht@physik.uni-wuerzburg.de)

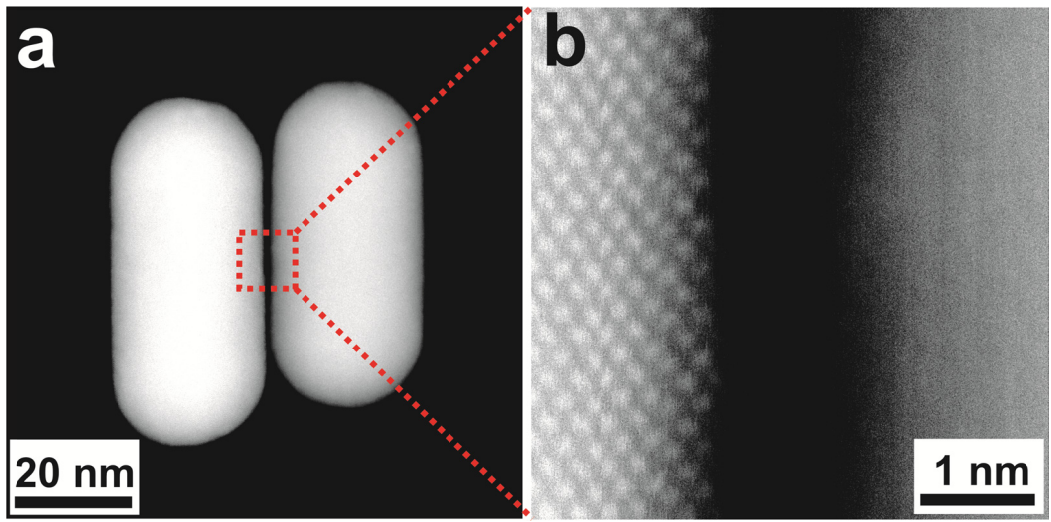

**Figure 1 | Characterization of a nanorod dimer by STEM.**

**a**, Dimer consisting of two side-by-side aligned gold nanorods with diameter of ~30 nm and length of ~65 nm, each. **b**, Zoom to the gap region indicated in **a** (dashed square). The atomic structure of the left nanorod is resolved. Detailed analysis of a number of gold nanorod dimers yields an average gap width of 1.3 nm (Supplementary).

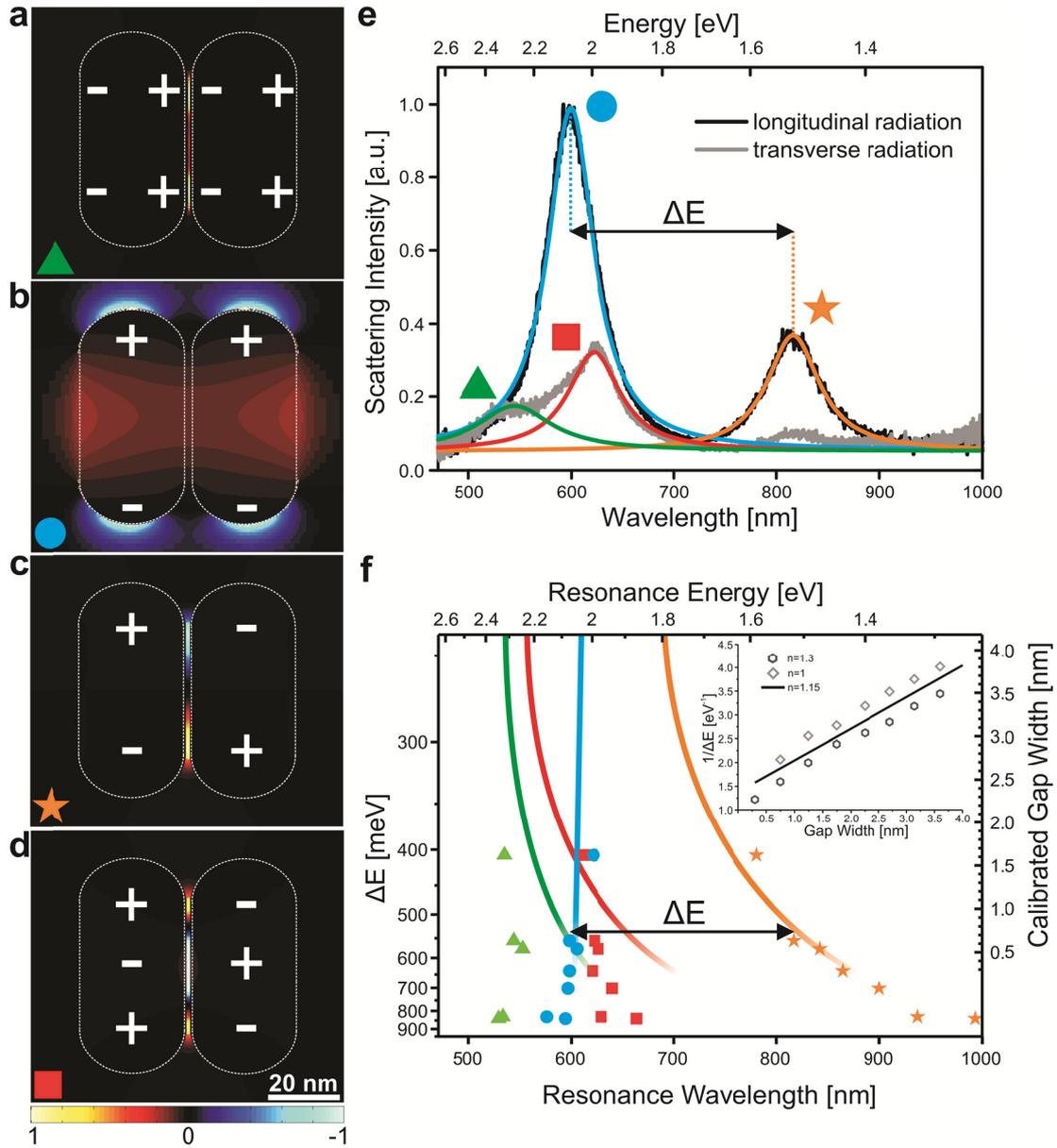

**Figure 2 | Numerical simulations and white-light scattering spectra of nanorod dimers.**

**a-d**, Simulated normalized near-field intensity distributions of the coupled transverse (green triangle), longitudinal (blue disk), 1st-order (orange star) and 2nd-order (red square) cavity eigenmodes, respectively, for a nanorod dimer. A phase-sensitive representation of the dominant field component, sgn(phase(E))*E$^2$, is used. **e**, Typical scattering spectrum of a nanorod dimer. Longitudinal (black) and transverse (grey) polarized detection with respect to the dimers long axis. Four resonances are observed to which Lorentzian line shapes are fitted [full lines, colored according to the assigned simulated eigenmodes **a-d**]. ΔE indicates the energy splitting between the coupled longitudinal and 1st-order cavity resonance. **f**, Resonance wavelength versus energy splitting between the coupled longitudinal and 1$^{st}$-order cavity resonance for the experimental data (symbols) and the simulated results (continuous lines). The gap width axis on the right is calibrated using the linear relationship (n=1.15) between inverse energy splitting (1/ ΔE) and gap width obtained by FDTD simulations (see inset).

**Supplementary Information**

**SEM images and size distribution of the investigated dimers**

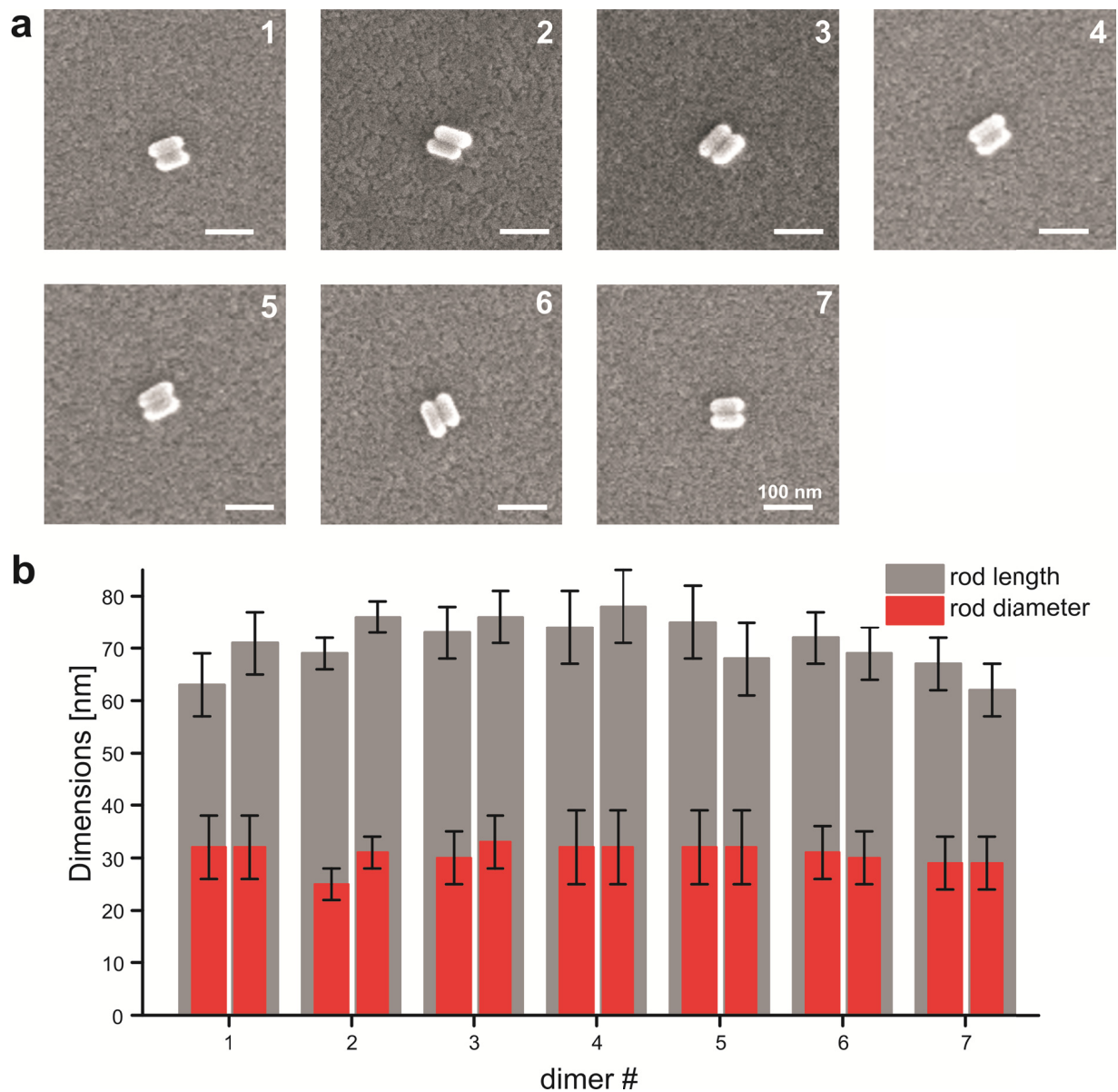

**Figure S1 | SEM images and size distribution of the investigated dimers.**

**a**, SEM images of all optically investigated dimers. **b**, For each dimer, the length and diameter of both nanorods constituting the dimer are plotted. The error of the measurement is given by the pixel size of the image.

**STEM study for the determination of the gap width of nanorod dimers**

To determine the gap width, STEM images of twelve parallel aligned rods were obtained. To prevent underestimation of the gap width due to parallax (Fig. S2a), for each pair a series of images was recorded for different sample angles, to obtain the micrograph with maximum geometric gap which was used for further analysis. As a next step the center region spanning about 20 nm along the length of the rods and 30 nm around the gap was selected (green box in Fig. S2b). Such an image basically consists of >200 cross sections of the gap that were all independently analyzed by nonlinear fitting. Intensity variations on the HAADF-STEM images arise due to changes in thickness and chemical composition and can therefore directly be used for gap width determination. As a fitting function the cross sectional thickness of a cylinder was used for both rods, with the radii, centers of the rod and the HAADF signal per thickness as free parameters. Exemplary cross sections at four different positions along the dimer and according fits are shown in Fig. S2c. A cylindrical curved surface was found to be appropriate for fitting the region around the gap. The results reflect the fact that the rods may not be perfectly parallel in every dimer; yet the angles between the rods were found to be smaller than 4 degrees in all cases. The mean gap value and its variance has been calculated from the >200 cross-sections for each rod (Fig. S2d). The obtained values for the gap were in good agreement with values obtained by simple visual inspection of the image. Ten out of twelve structures had gaps between 0.5 and ~2 nm, one pair (dimer #9) had a gap width too small to be determined and one outlier had a gap width of 3.5 nm. The average gap width for all structures was 1.3 nm.

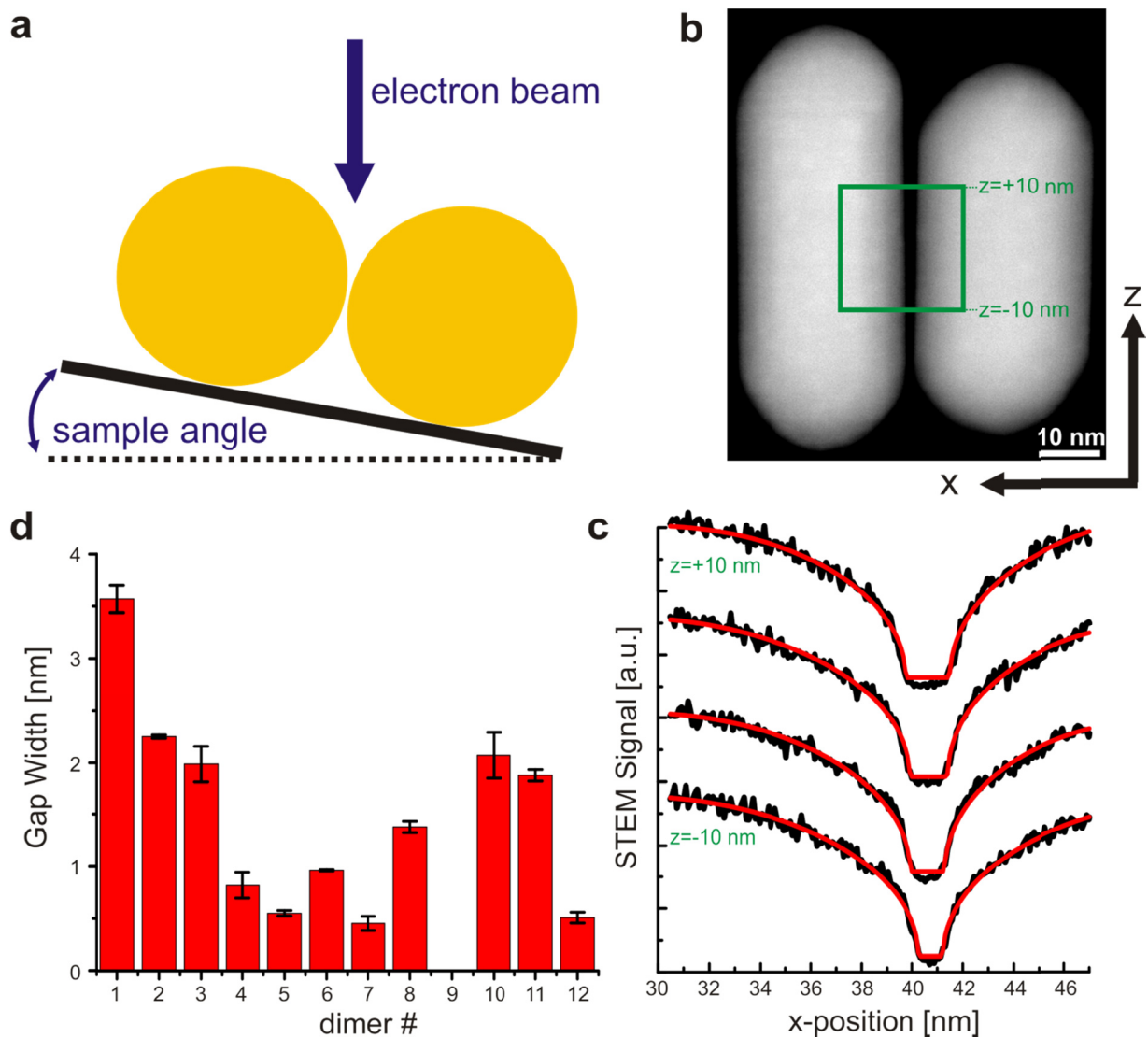

**Figure S2 | STEM study of the gap width of nanorod dimers.**

**a**, Sketch illustrating the importance of correct alignment for estimation of the gap width.

**b**, STEM image of one of the analyzed rods. The green box illustrates the region used to obtain the cross sections over the gap. **c**, Cross sections (solid black line) over the gap for different z-positions along the dimer. The solid red lines show the corresponding fit. **d**, Mean gap width and its variance for each dimer as obtained from >200 cross sections. Dimer #9 had a gap width too small to be determined.

**Simulated emission pattern of dimer modes**

The emission pattern reveals the radiation characteristics of a mode, i.e. the angular distribution and the polarization. According to the reciprocity theorem the emission pattern further reveals the most efficient excitation direction and polarization. In Fig. S3 we plot the emission pattern projected to a hemisphere below the structures. The dominating component in the far field determines the polarization and $E_x$/$E_z$ corresponds to transverse/longitudinal polarization respectively. The emission patterns were obtained from a 22x22 µm² large frequency monitor placed below the structure, by applying a far-field projection onto a hemisphere. A gap width of 10 nm and a mesh of 1 nm³ was used since emission patterns are rather insensitive to the gap width. For better visibility all patterns have been normalized to 1; typical ratios between two- and four-lobed patterns are on the order of 5. In accordance with symmetry arguments the far-field polarization is transverse for the coupled transverse and the 2$^{nd}$-order cavity resonance. A longitudinal polarization is observed for the coupled longitudinal and 1$^{st}$-order cavity resonance. Since the 1$^{st}$-order cavity resonance exhibits a quadrupolar-like charge distribution the radiation under angles normal to the substrate remains weak.

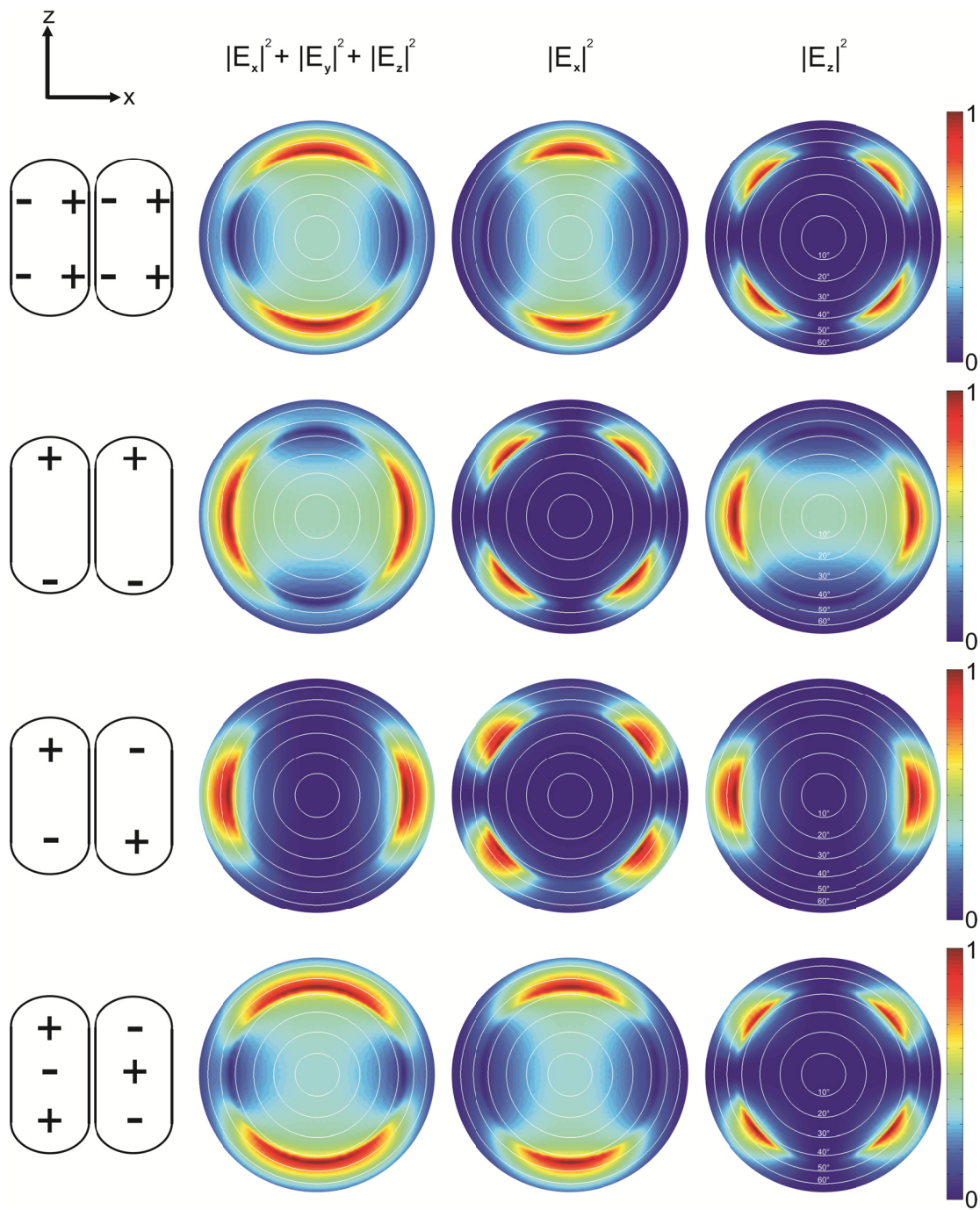

**Figure S3 | Simulated emission pattern of dimer modes.**

Simulated emission pattern for coupled transverse, coupled longitudinal, 1st-order cavity and 2nd-order cavity resonance (top to bottom). The far-field polarization of the modes is along the longitudinal direction (z) for the longitudinal and the 1st-order cavity resonance and for transverse and 2nd-order cavity resonance it is along the transverse direction (x). All emission patterns are normalized to 1 and have been projected to a hemisphere below the structure. White rings mark the angle of emission in steps of 10°.

**Analysis of quality factor and its dependence on slight structural asymmetries**

Various structures with different length and gaps were simulated and the quality factor of the longitudinal and both cavity modes was determined by fitting a Lorentzian line shape to the spectrum. In Fig. S4a we plot the quality factor versus resonance wavelength of longitudinal and both cavity modes for dimers with different lengths, gap widths and refractive indices in the gap. A specific wavelength dependence of the quality factor is observed for each mode, which does not seem to explicitly depend on details of the structural geometry but rather on the resonance wavelength of the mode. This indicates that for a given resonance the quality factor is mainly determined by its wavelength dependent ohmic losses.

The quality factor of the experimental data was obtained by fitting a Lorentzian line shape to each resonance of the measured spectra. In Fig. S4a we compare experimentally determined (solid symbols) and simulated (open symbols) quality factor for the coupled longitudinal and both cavity resonance. The data for the transverse resonance are not shown for clarity and since there is a large uncertainty in determination of the quality factor due to its poor excitation and weak radiation. Good agreement between simulation and experiments is observed for the coupled longitudinal resonance. For both cavity resonances a clear deviation between experiments and simulation is visible. For the $2^{nd}$-order cavity resonance a bigger deviation is observed for longer wavelengths, corresponding to smaller gaps. Such a trend is however not seen for the $1^{st}$-order cavity mode. Furthermore, the influence of slight structural asymmetries on the quality factor is studied and the corresponding field distributions are shown in Fig. S4b-d. The introduced asymmetries are a 5 nm offset between the rods (Fig. S4b), a tilt of one rod by two degrees (Fig. S4c) and decreasing the length of one rod by 5 nm (Fig. S4d). The shift of the resonance introduced by the asymmetries is less than 30 nm. The change in quality factor is small and the strongest effect is caused by tilting, where a 10% decrease is observed.

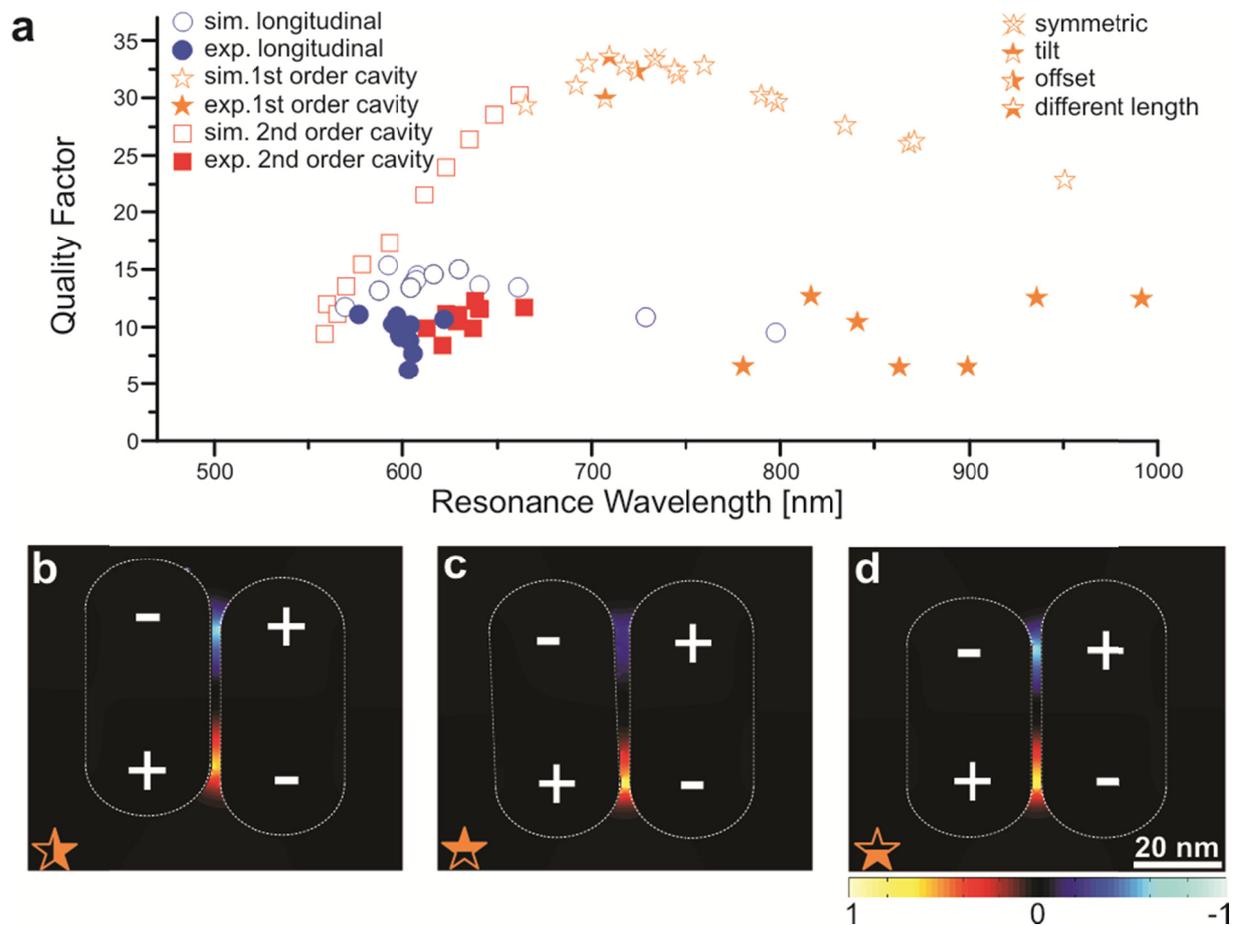

**Figure S4 | Analysis of quality factor and its dependence on slight structural asymmetries.**

**a**, Dependence of quality factor on resonance wavelength. Experimental and simulation data for the longitudinal and both cavity resonances are shown using solid and empty boxes respectively. To study the influence of slight asymmetries on the resonance wavelengths and quality factors, three types of asymmetries (half-filled orange stars) have been tested. The respective perfectly symmetric structure is indicated by the crossed-out empty orange star. **b-d**, Normalized near-field intensity distributions of the asymmetric structures. A phase-sensitive representation of the dominant field component, sgn(phase(E))*E2, is used. **b**, Offset of 5 nm between the two rods. **c**, Tilting of the left rod by two degrees. **d**, Decreasing the length of the left rod by 5 nm.

**Uncertainities due to meshing of small gaps in FDTD simulations**

Results obtained by the FDTD simulations software, used in this study, show very good agreement with an analytical model and no numerical artifacts occur for very small mesh sizes[S1].

We still investigated the magnitude of possible errors in the simulated resonance wavelength and quality factor introduced by the finite number of mesh cells across a 1 nm small gap of a nanorod dimer with diameter of 30 nm and length of 70 nm. The whole simulation area was meshed with a 0.25 nm mesh and a small mesh override region covers the area around the gap (see inset in Fig. S5). Resonance position and quality factor of the 2$^{nd}$-order cavity resonance, as determined by fitting a Lorentzian lineshape to the spectrum, are determined for 7 simulations in which the number of mesh cells in the gap is the only varied parameter.

Fig. S5 shows that resonance wavelength as well as quality factor vary less than 2% for different numbers of mesh cells. Convergence of the simulations for already a very low number of mesh cells is attributed to the fact that the gap modes for the present geometry exhibit very homogeneous field distributions across the gap with only small gradients of the electric field.

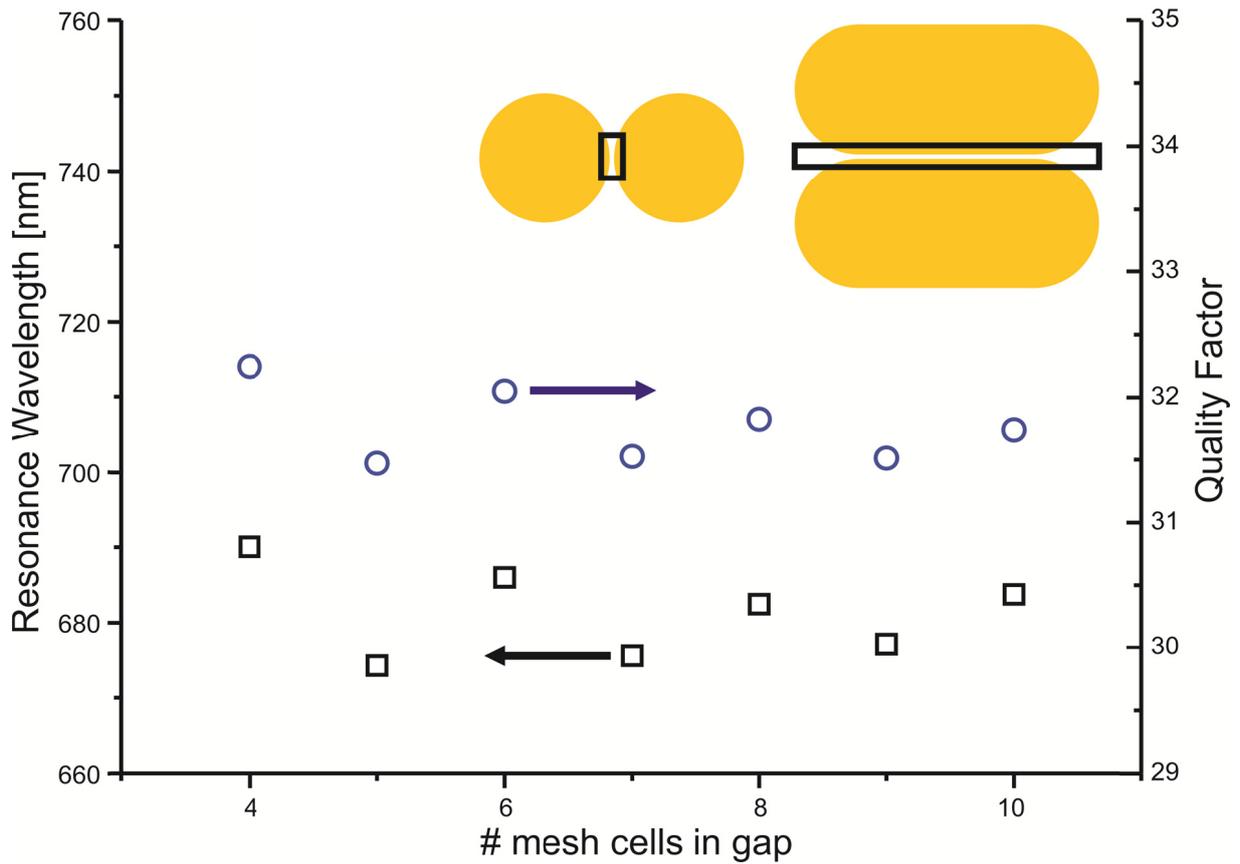

**Figure S5 | Uncertainties due to meshing of small gaps in FDTD simulations.**

Dependence of resonance wavelength (empty black boxes) and quality factor (empty blue circles) on number of mesh cells across the gap. Note the small wavelength and quality factor intervals used for plotting the minute changes in the resonance wavelength and quality factor.

**Supplementary References**

S1   FDTD Solutions Knowledge Base, Testing convergence. Retrieved from http://docs.lumerical.com/en/fdtd/user_guide_testing_convergence.html.